\documentclass{Interspeech}

\interspeechcameraready

\title{ReFlow-VC: Zero-shot Voice Conversion Based on Rectified Flow and Speaker Feature Optimization\thanks{$^{*}$Corresponding author}}

\author[affiliation={1}]{Pengyu}{Ren}
\author[affiliation={2}]{Wenhao}{Guan}
\author[affiliation={1}]{Kaidi}{Wang}
\author[affiliation={1}]{Peijie}{Chen}
\author[affiliation={*1}]{Qingyang}{Hong}
\author[affiliation={*2}]{Lin}{Li}

\affiliation{School of Informatics}{Xiamen University}{China}
\affiliation{School of Electronic Science and Engineering}{Xiamen University}{China}
\email{renpengyu@stu.xmu.edu.cn}

\usepackage{comment}

\begin{document}

\maketitle

\begin{abstract}
In recent years, diffusion-based generative models have demonstrated remarkable performance in speech conversion, including Denoising Diffusion Probabilistic Models (DDPM) and others. However, the advantages of these models come at the cost of requiring a large number of sampling steps. This limitation hinders their practical application in real-world scenarios. In this paper, we introduce ReFlow-VC, a novel high-fidelity speech conversion method based on rectified flow. Specifically, ReFlow-VC is an Ordinary Differential Equation (ODE) model that transforms a Gaussian distribution to the true Mel-spectrogram distribution along the most direct path. Furthermore, we propose a modeling approach that optimizes speaker features by utilizing both content and pitch information, allowing speaker features to reflect the properties of the current speech more accurately. Experimental results show that ReFlow-VC performs exceptionally well in small datasets and zero-shot scenarios.
\end{abstract}

\section{Introduction}

\begin{figure*}[t]
\centering
\includegraphics[width=1.0\textwidth]{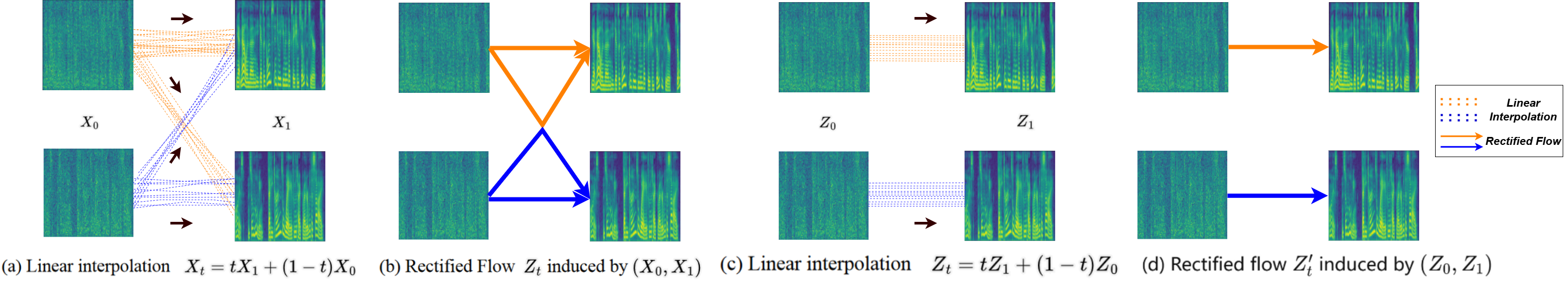} 
\caption{The diagram of the Rectified Flow model, with the meanings of the icons at the top.(a) Linear interpolation of data samples ($X_0$, $X_1$).(b) The rectified flow $Z_t$ induced by ($X_0$, $X_1$).(c) The linear interpolation of data samples ($Z_0$, $Z_1$) of rectified flow $Z_t$.(d) The rectified flow induced from ($Z_0$, $Z_1$), propagating along straight paths.}
\label{fig:1}
\end{figure*}

Zero-shot voice conversion (VC) aims to convert speech from any source speaker to the speech of any target speaker without changing the linguistic content. VC achieves this by decomposing the source speech into different components, including the speaker's timbre, linguistic content, and speaking style. Its applications span across various practical fields such as speech anonymization and audiobook production \cite{niminhua}. The core challenge of zero-shot VC lies in effectively modeling, disentangling, and utilizing various attributes of speech, including content and timbre.

Traditional zero-shot methods typically combine the linguistic content and speaking style of the source speaker with the timbre of the target speaker to generate the converted speech. In the groundbreaking Auto-VC model \cite{autovc}, speaker embeddings from a pre-trained speaker verification network are used as conditional inputs. Other models have been improved upon this by enriching the conditional inputs with additional speech features, such as pitch and loudness, or by jointly training voice conversion and speaker embedding networks. Additionally, several studies have utilized attention mechanisms to better integrate the features of reference speech into the source speech, thus enhancing the performance of the decoder \cite{333}. However, due to the complexity of speech signals \cite{444} and the limitations in modeling timbre and content \cite{555}, these methods still leave significant room for performance improvement.

Most traditional voice conversion models are based on autoencoder architectures, such as Auto-VC, VAE-VC \cite{vaevc}, and CycleGAN-VC \cite{CycleGANVC}, or generative adversarial networks (GANs) \cite{gan}, like CycleGAN-VC \cite{CycleGANVC}, CycleGAN-VC2 \cite{CycleGANVC2}, and StarGAN-VC \cite{StarGAN-VC}. However, relatively few voice conversion models are based on flow models.

Notable voice conversion models, such as Free-VC \cite{freevc}, Auto-VC, and Diff-VC \cite{diffvc}, have achieved significant success. However, there remains room for exploration in terms of model architecture and methodology. Among them, diffusion models, including denoising diffusion probabilistic models (DDPM) and score-based generative models, have gained widespread attention due to their potential to generate high-quality samples. A major drawback of diffusion models, however, is that generating satisfactory samples requires multiple iterations \cite{reflowtts}. To address this, several voice conversion methods have been proposed. Diff-VC gradually converts noise into Mel spectrograms by constructing a stochastic differential equation (SDE) and solving the reverse SDE with a numerical ordinary differential equation (ODE) solver. While it generates high-quality audio, the large number of iterations in the reverse process affects inference speed. DDDM-VC addresses the issue of multiple iterations required for generating high-quality samples by introducing a decoupled denoising diffusion model and a prior mixing strategy \cite{DDDMVC}. In this paper, we present ReFlow-VC, a voice conversion model based on a modified flow model. Our proposed method achieves outstanding voice conversion results, and the contributions of this paper are as follows:
\begin{itemize}
    \item We propose ReFlow-VC, the first voice conversion acoustic model based on a Rectified Flow Model. Specifically, the ReFlow-VC model is an ordinary differential equation (ODE) model that transforms a Gaussian distribution into the true mel-spectrogram distribution via a direct path as much as possible, and is trained using a simple unconstrained least squares optimization procedure \cite{reflowtts}.
    \item We propose a modeling approach that optimizes speaker features using both content and pitch. Through cross-attention and gated fusion, it effectively integrates multiple input features (such as speaker characteristics, content information, pitch, etc.), thereby enhancing the model's expressive capability. This enables fine-grained control over the target speaker’s attributes, making the voice conversion task more precise.
\end{itemize}

\section{Rectified Flow Model}
\subsection{Rectified Flow}
The rectified flow model is an Ordinary Differential Equation (ODE) model designed to transform a distribution $\pi_0$ (standard Gaussian) to $\pi_1$ (the ground truth distribution) via straight-line paths. Given samples $X_0 \sim \pi_0$ and $X_1 \sim \pi_1$, the rectified flow corresponds to an ODE defined as \cite{flow}:
\begin{equation}
    dZ_t = v(Z_t, t) \, dt,
\end{equation}where $Z_0$ is from $\pi_0$, and the transformation follows the distribution $\pi_1$. Here, $\upsilon$ represents the drift force of the ODE, designed to align the flow with the direction ($X_1$-$X_0$) between the two distributions. The flow is learned by minimizing a least squares regression problem:

\begin{equation}
    \underset{\upsilon}{\min} \int_0^1 \lVert (X_1 - X_0) - v(X_t, t) \rVert^2 \, dt,
\end{equation}where $X_t$ is the linear interpolation between $X_0$ and 
$X_1$, defined as:

\begin{equation}
    X_t = t X_1 + (1 - t) X_0.
\end{equation}

While the naive evolution of $X_t$ follows a non-causal path $dX_t = (X_1 - X_0) \, dt$, the rectified flow causalizes the path by adjusting $\upsilon$ based on $(X_1 - X_0)$, ensuring that the trajectories do not cross at any point, preserving the uniqueness of the solution. The rectified flow thus avoids non-causal intersections, as shown in Figure \ref{fig:1}, ensuring a well-defined, non-crossing path.During training, the objective is to learn the drift force $\upsilon$ by minimizing:

\begin{equation}
    \hat{\theta} = \arg \min_{\theta} \mathbb{E}\left[\lVert (X_1 - X_0) - v(X_t, t) \rVert^2\right],
\end{equation}where t$\sim$Uniform([0,1]). After training, the learned model is used to transform $X_0$ to $X_1$ by solving the ODE $dZ_t=\hat{v}(Z_t, t) \, dt$. This procedure can be recursively applied, forming a sequence of transformations $Z^{'} = \text{ReFlow}(Z_0, Z_1)$, leading to improved transport efficiency and more linear flow trajectories. This recursive process helps reduce time-discretization errors and is computationally advantageous when simulating flows.

\subsection{Rectified Flow Model for VC}

ReFlow-VC converts the noise distribution to a Mel spectrogram distribution conditioned on time t and the speaker condition features c after feature fusion. We define $\pi_0$ as the standard Gaussian distribution and $\pi_1$ as the ground truth Mel-spectrogram data distribution, with $X_0 \sim \pi_0$ and $X_1 \sim \pi_1$. The training objective of ReFlow-VC is as follows:

\begin{equation}
    L_\theta = \mathbb{E}\left[\lVert (X_1 - X_0) - v_\theta(X_t, t, c) \rVert^2\right],
\end{equation}where $t \in \text{Uniform}([0, 1])$ and $X_t = t X_1 + (1 - t) X_0$. ReFlow-VC does not require any auxiliary losses, except for the L2 loss function between the output of the model $v_\theta$ and $(X_1-X_0)$. During inference, we directly solve the ODE starting from $Z_0 \sim \pi_0$ conditioned on the speaker feature c and based on the model $v_\theta$. For high-fidelity generation, we can use the RK45 ODE solver. For one-step generation, we can directly use the Euler ODE solver for competitive performance \cite{reflowtts}.
Furthermore, the recursive rectified flow procedure can also be applied to VC, constructing a second ReFlow-VC, referred to as 2-ReFlow-VC. The 2-ReFlow-VC is simply retraining the rectified flow model using the samples generated by ReFlow-VC.

\section{ReFlow-VC Model Architecture}

\begin{figure*}[t]
\centering
\includegraphics[width=1.0\textwidth]{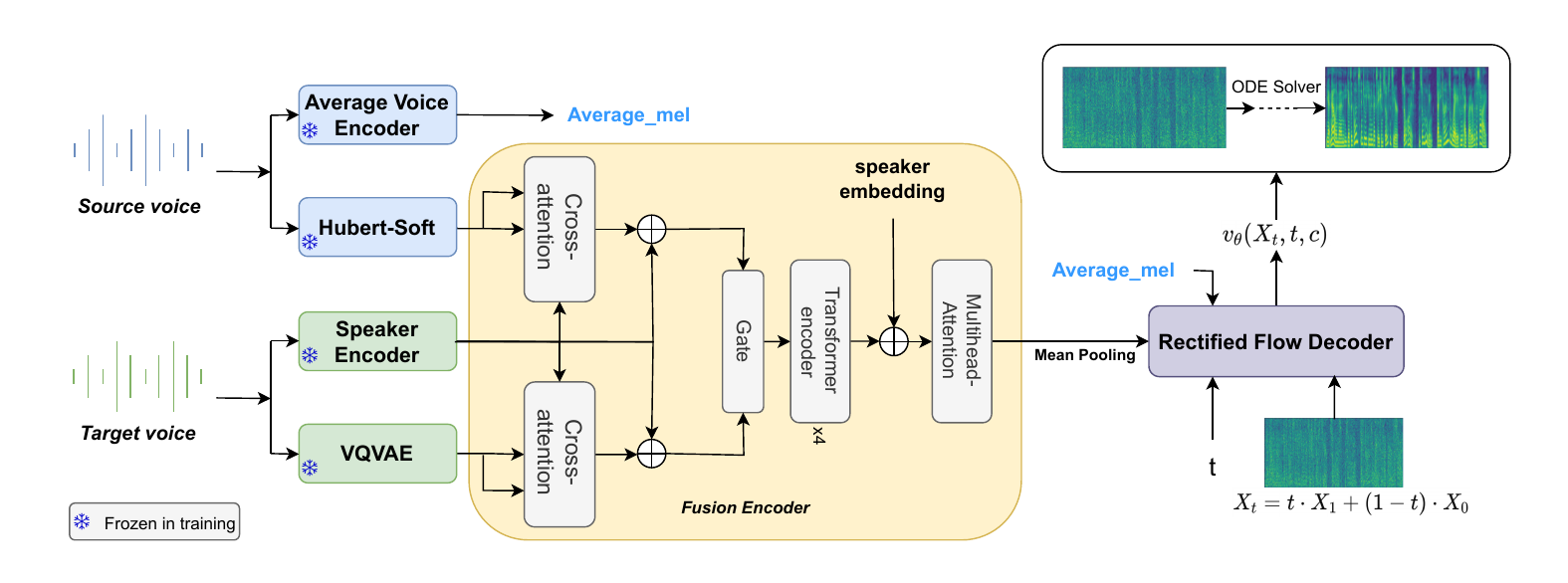} 
\caption{An illustration of ReFlow-VC.}
\label{fig:fig1}
\end{figure*}

\subsection{Encoder}
The encoder consists of an average voice encoder, Hubert-Soft \cite{softvc}, a speaker encoder, VQ-VAE, and a feature fusion module. We chose the average phoneme-level MEL features as the speaker-independent speech representation, similar to Diff-VC \cite{diffvc}.By using the average voice encoder, the source audio is transformed into the average speaker's mel, which is referred to as Average\_mel in Figure \ref{fig:fig1}. Additionally, we used Hubert-Soft from Soft-VC to extract continuous content features. By modeling uncertainty, Hubert-Soft captures more content information, which improves the clarity and naturalness of the converted speech. Following the work of Polyak et al \cite{pitch}, we used the YAPPT algorithm to extract pitch (F0) from the audio, encoding speaker-independent pitch information. The F0 of each sample is normalized for each speaker to obtain speaker-independent pitch information, and the VQ-VAE is used to extract vector quantized pitch representations \cite{pitch}. For a fair comparison, during inference, we normalize the F0 for each sentence instead of for each speaker.

Essentially, zero-shot voice conversion is a challenging task, as it requires the model to generalize effectively to any unseen speaker without additional training or fine-tuning. This places high demands on the model's ability to capture timbre \cite{zsvc}. To enhance the overall timbre modeling capability of the proposed method, we introduce a feature fusion module. Through multiple attention mechanisms, gated fusion, and iterative self-attention strategies, the speaker features can be precisely adjusted and optimized across various speech attributes \cite{123}. Specifically, the model can dynamically adjust speaker features, flexibly modify them using content and pitch information, and enhance the expressiveness of speaker characteristics. By leveraging cross-attention and gating mechanisms, it captures and generates personalized speaker traits with greater accuracy. Additionally, the model improves the accuracy and naturalness of voice conversion, ensuring clear and natural speech across different speakers through the use of multiple attention and self-attention mechanisms. These strategies significantly improve the expressiveness of the speaker features, resulting in more precise and natural voice conversion outcomes.

Specifically, the fusion encoder consists of multiple modules designed to enhance performance and expressive power. The pitch conv projection layer transforms the input pitch features from a dimension of 1 to 256. Then, through two layers of cross-attention, each layer receives an input of 256 dimensions and outputs 256 dimensions to facilitate mutual attention between different features. Next, the gated fusion module processes the 256-dimensional input to enhance the information fusion effect. The self-attention mechanism iteratively refines the 256-dimensional input's self-attention performance, gradually focusing on the key parts of the input. Additionally, the model employs a multihead-attention mechanism with both input and output dimensions of 256 and 8 attention heads (with a dimension of 32 per head), further enhancing the model's ability to learn and focus on various features. These modules work together to improve the model's ability to process speech features and speaker conditioning information.

\subsection{Decoder}
The decoder architecture is based on U-Net \cite{unet} and is the same as in GradTTS \cite{gradtts} but with four times more channels to better capture the full range of human voices. The speaker conditioning network $g_t(Y)$ consists of 2D convolutions and MLPs(multilayer perceptron). Its output is a 128-dimensional vector, which is broadcast-concatenated with the concatenation of ${\hat{X}}_t$  and $\bar{X}$ as additional 128 channels.
\section{Experiments And Results}
\subsection{Experimental Setup}

\begin{table*}[ht]
\centering
\begin{tabular}{p{4cm}|>{\centering\arraybackslash}p{0.6cm}|>{\centering\arraybackslash}p{2.3cm}|>{\centering\arraybackslash}p{2.3cm}|>{\centering\arraybackslash}p{1.4cm}|>{\centering\arraybackslash}p{1.4cm}|>{\centering\arraybackslash}p{1.4cm}}
\hline
\textbf{Method} & \textbf{iter.} & \textbf{NMOS} (\(\uparrow\)) & \textbf{SMOS} (\(\uparrow\)) & \textbf{CER} (\(\downarrow\)) & \textbf{WER}(\(\downarrow\))& \textbf{SECS} (\(\uparrow\)) \\
\hline
\textbf{GT} & - & 3.89 $\pm$ 0.05 & 3.47 $\pm$ 0.05 & 0.52 & 1.79 & -   \\

\textbf{GT (Mel + Vocoder)} & - & 3.86 $\pm$ 0.05 & 3.34 $\pm$ 0.05 & 0.59 & 2.16 & 0.987  \\
\hline
\textbf{AutoVC} & - & 3.59 $\pm$ 0.05 & 2.41 $\pm$ 0.04 & 5.47 & 8.69 & 0.746 \\
\hline
\textbf{Free-VC} & - & 3.76 $\pm$ 0.05 & 2.78 $\pm$ 0.05 & \textbf{1.69} & 4.77  & 0.745 \\
\hline
\textbf{Diff-VC } & 1 & 2.13 $\pm$ 0.05 & 1.78 $\pm$ 0.04 & 7.49 & 13.05 & 0.617  \\

\textbf{Diff-VC } & 30 & 3.72 $\pm$ 0.05 & 2.69 $\pm$ 0.05 & 7.58 & 13.80 & 0.766  \\

\textbf{Diff-VC} & 1000 & 3.75 $\pm$ 0.05 & 2.78 $\pm$ 0.05 & 7.64 & 13.92  & 0.781 \\
\hline
\textbf{NReFlow-VC} & 1 & 3.69 $\pm$ 0.05 & 2.67 $\pm$ 0.05 & 4.35 & 7.60 & 0.751  \\

\textbf{NReFlow-VC} & 30 & 3.74 $\pm$ 0.05 & 2.70 $\pm$ 0.05 & 4.45 & 7.71 & 0.759  \\

\textbf{NReFlow-VC(RK45 solver)} & 146 & 3.75 $\pm$ 0.05 & 2.70 $\pm$ 0.05 & 4.75 & 7.79 & 0.761  \\
\hline
\textbf{ReFlow-VC} & 1 & 3.72 $\pm$ 0.05 & 2.74 $\pm$ 0.05 & 1.83 & \textbf{4.54} & 0.830  \\

\textbf{ReFlow-VC} & 30 & 3.74 $\pm$ 0.05 & 2.76 $\pm$ 0.05 & 1.96 & 4.59 & \textbf{0.845}  \\

\textbf{ReFlow-VC(RK45 solver)} & 512 & \textbf{3.78} $\pm$ \textbf{0.05} & \textbf{2.81} $\pm$ \textbf{0.05} & 2.12 & 4.84  & 0.843 \\
\hline
\end{tabular}
\caption{Evaluation results for VC.}
\label{table:t1}
\end{table*}

\begin{table}
\centering
\begin{tabular}{p{4cm}|>{\centering\arraybackslash}p{1.4cm}|>{\centering\arraybackslash}p{1.4cm}}
\hline
\textbf{Method} & \textbf{iter.} & \textbf{SECS} (\(\uparrow\)) \\
\hline
\textbf{ReFlow-VC}    & 1 & 0.830  \\

\textbf{ReFlow-VC}  & 30 & \textbf{0.845}  \\

\textbf{ReFlow-VC(RK45 solver)}& 512 & 0.843 \\
\hline
\textbf{2-ReFlow-VC} & 1  & 0.832 \\

\textbf{2-ReFlow-VC }  & 30 & 0.841  \\

\textbf{2-ReFlow-VC(RK45 solver) }  & 396 & 0.842  \\
\hline
\end{tabular}
\caption{Evaluation results for ReFlow-VC and 2-ReFlow-VC.}
\label{table:t2}
\end{table}

\subsubsection{Dataset}
We trained the proposed ReFlow-VC model using a subset of the LibriTTS dataset, which was randomly sampled from the full LibriTTS dataset to facilitate model training on a small dataset \cite{libritts}. This subset consists of 26,580 speech samples with a total duration of 38.13 hours. The dataset was randomly split into a training set (26,000 samples), a validation set (80 samples), and a test set (500 samples). We extracted 80-dimensional mel spectrograms with a frame size of 1024 and a hop size of 256. ReFlow-VC was trained for 200K iterations on a single NVIDIA 2080Ti GPU using the Adam optimizer \cite{adam}. We used the pre-trained HiFi-GAN \cite{hifigan} as the neural vocoder, which is responsible for converting the mel spectrograms into raw waveforms.
\subsubsection{Evaluation Metrics}
We conducted a comprehensive evaluation, including both objective and subjective metrics, to assess the sample quality (NMOS, SMOS, WER, CER, and SECS) as well as the model's inference speed. We calculated the Character Error Rate (CER) and Word Error Rate (WER) using Whisper \cite{cao2012whisper}, a publicly available Automatic Speech Recognition (ASR) model that is large-scale, multilingual, and multitask-supervised, for content consistency measurement. We evaluated the speech naturalness and speaker similarity through Mean Opinion Scores (MOS). For speech naturalness (NMOS), at least 20 listeners rated each source and converted speech sample on a scale of 1 to 5. For speaker similarity (SMOS), at least 20 listeners rated the target and converted speech samples on a scale of 1 to 4. Additionally, we performed an extra similarity measurement using the speaker encoder’s cosine similarity (SECS).

\subsubsection{Comparative Models}
We compared the performance of the samples generated by ReFlow-VC with the following systems using the six metrics mentioned above: 1) Diff-VC\footnote{https://github.com/huawei-noah/Speech-Backbones/tree/main/DiffVC}; 2) Auto-VC\footnote{https://github.com/auspicious3000/autovc}; 3) Free-VC\footnote{https://github.com/OlaWod/FreeVC}. Note that the Diff-VC model was re-trained by us on the same small dataset as ReFlow-VC, and to achieve better performance, the time step T of Diff-VC was set to 1000. Additionally, we conducted an ablation experiment, naming the version of ReFlow-VC without the feature fusion module as NReFlow-VC for comparative testing.

\subsection{Results and Analysis}
The evaluation results of VC are shown in Table \ref{table:t1}. In terms of audio quality, our proposed ReFlow-VC model, using the RK45 ODE solver for inference, achieves the highest SMOS, NMOS, and best SECS scores among all methods. This demonstrates the superior performance of our ReFlow-VC in modeling data distributions. We observe that compared to Diff-VC, the one-step sampling performance of both ReFlow-VC and NReFlow-VC is quite impressive, especially the one-step sampling performance of ReFlow-VC, which is on par with Diff-VC's 30-step sampling performance. Additionally, when comparing ReFlow-VC and NReFlow-VC, we find that our proposed cross-attention and gated fusion module, which models speaker characteristics using content and pitch features, plays a crucial role in improving the speaker naturalness and similarity of the converted speech, yielding significant improvements. Under equal sampling, ReFlow-VC's SECS score clearly outperforms both NReFlow-VC and Diff-VC. However, we also note that there is still room for improvement in CER and WER data.

In addition, we used a 9.2 second female audio as the source speaker's audio and a 4.7 second male audio as the target speaker's audio to test the sampling duration. This duration only accounts for the time spent during the sampling phase, excluding the time taken for extracting pitch and content features. As shown in Table \ref{table:t3}, with the same number of sampling steps, Rectified Flow achieves faster sampling speed. However, due to the additional feature fusion module in ReFlow-VC compared to NReFlow-VC, the sampling time is longer. Additionally, ReFlow-VC requires extra time for extracting pitch and content features. In the above example, extracting content features with Hubert-Soft took 0.0263 seconds, and extracting pitch features took 1.0468 seconds. However, considering the overall speech conversion results, we believe the additional time consumption is worthwhile.

We also conducted experiments with 2-ReFlow-VC, as shown in Table \ref{table:t2}. The results indicate that although 2-ReFlow-VC performs similarly to ReFlow-VC, it performs slightly better when using the Euler ODE solver (1-step). However, when using the Euler ODE solver (30-steps) and the RK45 solver, ReFlow-VC performs slightly better than 2-ReFlow-VC. Despite these differences, they are not significant. This further proves that the recursive rectified flow model is more intuitive and easier to implement numerically. By exploring the processes of ReFlow-VC and 2-ReFlow-VC, we highlight the robustness of our proposed conditional rectified flow model.

\begin{table}
\centering
\begin{tabular}{p{4cm}|>{\centering\arraybackslash}p{1.4cm}|>{\centering\arraybackslash}p{1.4cm}}
\hline
\textbf{Method} & \textbf{iter.} & \textbf{Time(s)} \\
\hline
\textbf{Diff-VC}    & 1 & 0.1188  \\
\textbf{Diff-VC}    & 30 & 4.6773  \\
\textbf{Diff-VC}    & 1000 & 167.4375  \\
\hline
\textbf{NReFlow-VC}    & 1 & \textbf{0.0893}  \\

\textbf{NReFlow-VC}  & 30 & \textbf{4.4197}  \\

\textbf{NReFlow-VC(RK45 solver)}& 140 & \textbf{22.2049} \\
\hline
\textbf{ReFlow-VC} & 1  & 0.0998 \\

\textbf{ReFlow-VC }  & 30 & 4.452  \\

\textbf{ReFlow-VC(RK45 solver) }  & 554 & 88.79923 \\
\hline
\end{tabular}
\caption{The sampling duration of Diff-VC, NReFlow-VC, and ReFlow-VC.}
\label{table:t3}
\end{table}

\section{Conclusions}

In this paper, we propose a simple yet efficient ReFlow-VC, which effectively completes the voice conversion task by leveraging rectified flow and feature fusion techniques. ReFlow-VC can use the RK45 ODE solver for sampling, generating speech samples with optimal audio quality. Furthermore, due to the excellent one-step sampling capability and outstanding performance of our proposed ReFlow-VC, it significantly enhances its usability in real-world scenarios. The audio samples are publicly available at: \href{https://copyabcs.github.io/reflowvc-demo/}{https://copyabcs.github.io/reflowvc-demo/.}

\section{Acknowledgements}

This work was supported in part by the National Natural Science Foundation of China under Grants 62276220 and 62371407 and the Innovation of Policing Science and Technology, Fujian province (Grant number: 2024Y0068)

\newpage

\bibliographystyle{IEEEtran}
\bibliography{mybib}

\end{document}